\documentclass[fleqn,twoside]{article}
\usepackage{espcrc2}

\usepackage{graphicx}
\usepackage[figuresright]{rotating}

\newcommand{\be}{\begin{equation}} \newcommand{\ee}{\end{equation}}
\newcommand{\ba}{\begin{eqnarray}} \newcommand{\ea}{\end{eqnarray}}

\newcommand{\AmS}{{\protect\the\textfont2
  A\kern-.1667em\lower.5ex\hbox{M}\kern-.125emS}}

\title{ Superscaling and Charge-changing Neutrino Cross Sections}
\author{M.B. Barbaro,
	J.E. Amaro,
	J.A. Caballero,
	T.W. Donnelly,
	A. Molinari,
	I. Sick
        \address{Dip. di Fisica Teorica, Univ. di Torino and INFN, 
        Via P. Giuria 1, I10125 Turin, Italy,
        \\
	Dep. de F\'\i sica Moderna,
	  Univ. de Granada, 18071 Granada, Spain,
	\\
	Dep. de F\'\i sica At\'omica, Molecular y Nuclear,
	  Univ. de Sevilla, Apdo. 1065, 41080 Sevilla, Spain,
	\\
	CTP, LNS 
	  and Dep. of Physics, Massachusetts Institute of Technology,
	  Cambridge, MA 02139, USA,
	\\
	Dep. f\"ur Physik und Astronomie,
	  Univ. Basel, CH-4056 Basel, Switzerland } 
}
       
\begin{document}

%\begin{abstract}
%  The superscaling function extracted from inclusive electron scattering data
%is used to predict high energy charge-changing neutrino 
%cross sections in the quasi-elastic and $\Delta$ regions.
%\vspace{1pc}
%\end{abstract}

%\vspace{-1.cm}
\maketitle

\vspace{-2.cm}

High-quality predictions for neutrino-nucleus cross sections are needed
for use in on-going experimental studies of neutrino oscillations at 
GeV energies, where a fully relativistic
treatment of the neutrino-nucleus scattering is mandatory, but hard to
achieve.  
Of course any reliable calculation for neutrino scattering should
first be tested against electron scattering in the same kinematical 
conditions. 
In fact, while relativistic modeling of the nuclear dynamics is
capable of getting the basic size and shape of the inclusive cross 
section, it can hardly account for important details of the response. 
Specifically, 
the data display a long tail which is largely absent in most modeling,
possibly due to the absence of classes of short-range correlation effects
and to the treatment of final-state interactions.

In order to overcome these difficulties, instead of using a specific 
nuclear model, we propose here a method for extracting informations 
on the nuclear dynamics from electron scattering experiments and use 
them to predict the neutrino-nucleus cross section. 

The approach is based on the superscaling behavior of the electron-nucleus 
cross section in both the quasi-elastic (QE) and $\Delta$ peaks. 
Detailed studies of superscaling can be found in 
Refs.~\cite{scaling1,scaling2}; here we simply recall
the basic idea. By dividing the experimental inclusive cross section for
various momentum transfers $q$ and nuclei (or, equivalently,
Fermi momentum $k_F$) by an appropriate single-nucleon cross section, 
a reduced cross section $f$, called the superscaling function, 
is obtained, which embodies the nuclear dynamics. This
can be plotted versus a well-chosen scaling variable:
if no dependence of $f$ upon $q$ is observed, 
we call this behavior  scaling of the 
first kind, whereas if no dependence occurs on the specific nucleus,
we call it scaling of the second kind; if both types of
scaling behavior are found we say that superscaling occurs.

Several choices for the appropriate scaling variable have been proposed 
in the literature. Here we choose that which naturally emerges from the
relativistic Fermi gas (RFG) model. In this framework the
nuclear response functions in both the quasi-elastic ($X=QE$) 
and $\Delta$-resonance ($X=\Delta$) regions have the general structure
\be
R_X =\frac{{\cal N} m_N}{q k_F} 
\left[ R_X \right]^{s.n.} f_{RFG}^{X}(\psi_{X} )
\label{R}
\ee
where 
${\cal N}$ is the appropriate nucleon number and
$\left[ R_X \right]^{s.n.}$ is the single-nucleon response function.
The RFG superscaling function 
\be
f^{QE}_{RFG}(\psi) =
f^{\Delta}_{RFG}(\psi) 
=\frac{3}{4}(1-\psi^2)\theta(1-\psi^2)
%\nonumber
\label{f}
\ee
is independent of $q$ and $k_F$, being a function of a specific combination
of $q$, $\omega$ and $k_F$, the scaling variable, defined as
%\begin{eqnarray}
\be
\psi _{X}=
\pm 
\sqrt{
%\left[ 
%\left( 
%\epsilon_{0,X}
\frac{1}{T_F}\left(\frac{q}{2} 
\sqrt{\rho_X^{2}+1/\tau }-\frac{\omega}{2} \rho_X-m_N\right),
%\right) 
%\right] ^{1/2} , 
%\nonumber\\
}
\label{psi}
%\end{eqnarray}
\ee
where $T_F$ is the Fermi kinetic energy, $4 m_N^2\tau=q^2-\omega^2$ and
$\rho_X=1+(m_X^2-m_N^2)/(q^2-\omega^2)$ ($m_{QE}$ being the nucleon mass).
The sign in Eq.~(\ref{psi}) is $+$ ($-$) if the energy transfer $\omega$
is higher (lower) than the one corresponding to the maximum of the peak.
Moreover, a small energy shift is introduced, implying $\psi\to\psi^\prime$,  
in order to force the maximum of the response to occur for 
$\psi^\prime_X=0$.

At sufficiently high energies the electron scattering data
exhibit both types of scaling behavior in the quasi-elastic region.
For specific nuclei the first-kind scaling is quite good at excitation
energies below the QE peak (the so-called scaling region): this is
the familiar $y$-scaling behavior. At energies above the peak, where
nucleon resonances (especially the $\Delta$) are important, this type of
scaling is broken.
The scaling violations apparently reside in the transverse
response, but not in the longitudinal, which appears to superscale.  In
fact, this is not unexpected, since MEC and inelastic contributions, which do 
not scale, are predominantly transverse in the kinematic regions of
interest. 
On the other hand 
scaling of the second kind works very well in the scaling region
and, even in the resonance region, is only violated at roughly the 20\% level.

The approach taken in \cite{scaling2}
therefore has been to use the experimental longitudinal responses to 
define the superscaling function. 
Reliable separations of data into their longitudinal and transverse
contributions for $A>4$ are available only for a few
nuclei; all of these response functions have been used to
extract the ``universal'' quasi-elastic response function $f^{QE}$ and
to obtain a parameterization by a simple function. Fig.~1 (upper panel) 
shows $f^{QE}(\psi'_{QE})$ averaged over the nuclei employed, together with the
corresponding fit, plotted against the scaling variable
$\psi'_{QE}$. 

Note that $f^{QE}(\psi'_{QE})$ 
has a somewhat asymmetric shape and a tail that extends towards positive 
values of $\psi'_{QE}$. In contrast, the RFG superscaling function
(\ref{psi}) is symmetric, is limited strictly to the region 
$-1\leq \psi^\prime_{QE}\leq +1$ and has a maximum value of 3/4, while the 
empirical superscaling function reaches only about 0.6. The microscopical
origin of this asymmetric shape has been the object of recent 
investigations in the context of the relativistic impulse
approximation~\cite{RIA}.
\begin{figure}[hbt]
\begin{center}
\includegraphics[scale=0.4,clip,angle=0]{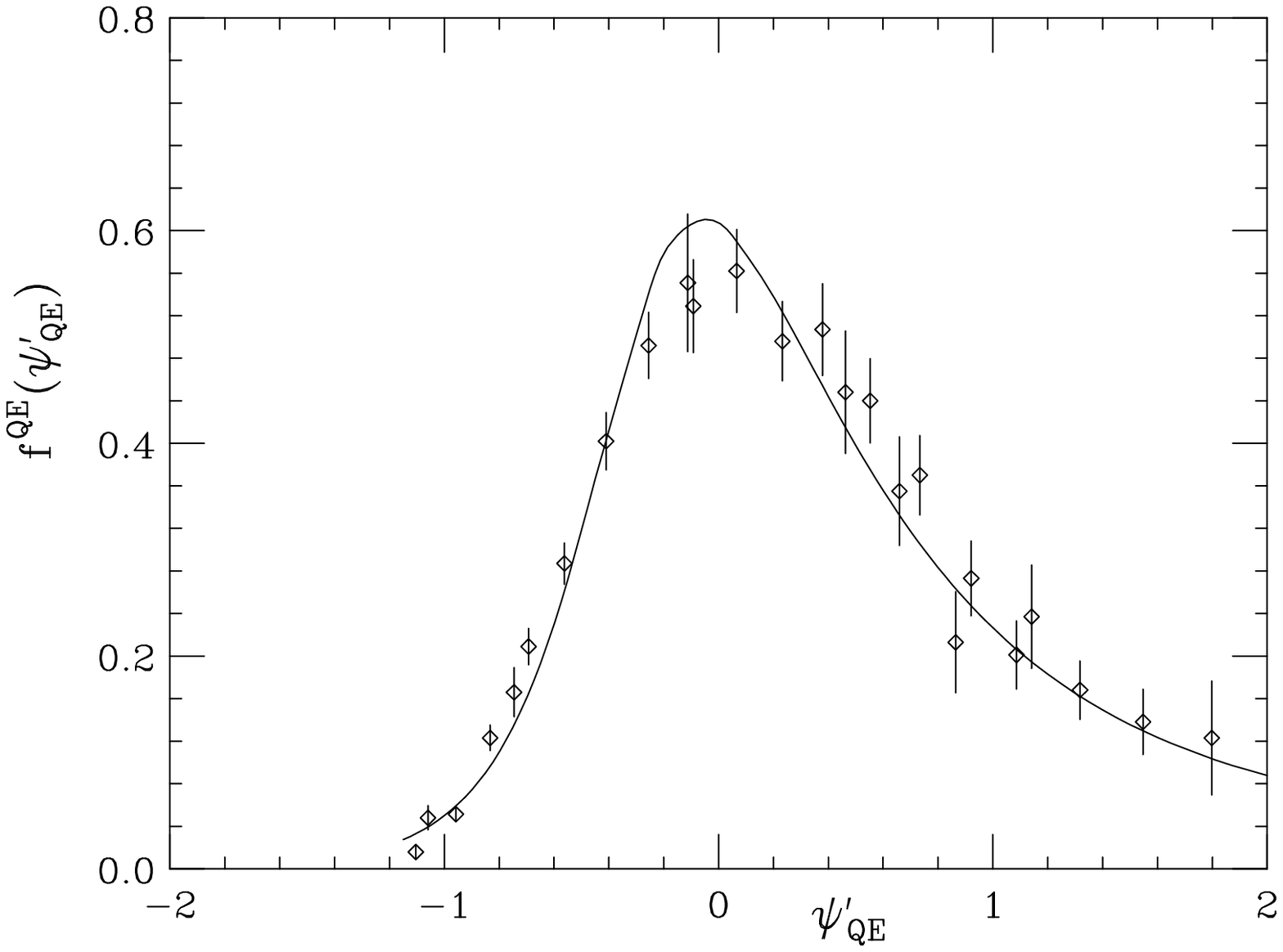}
\includegraphics[scale=0.38,clip,angle=0]{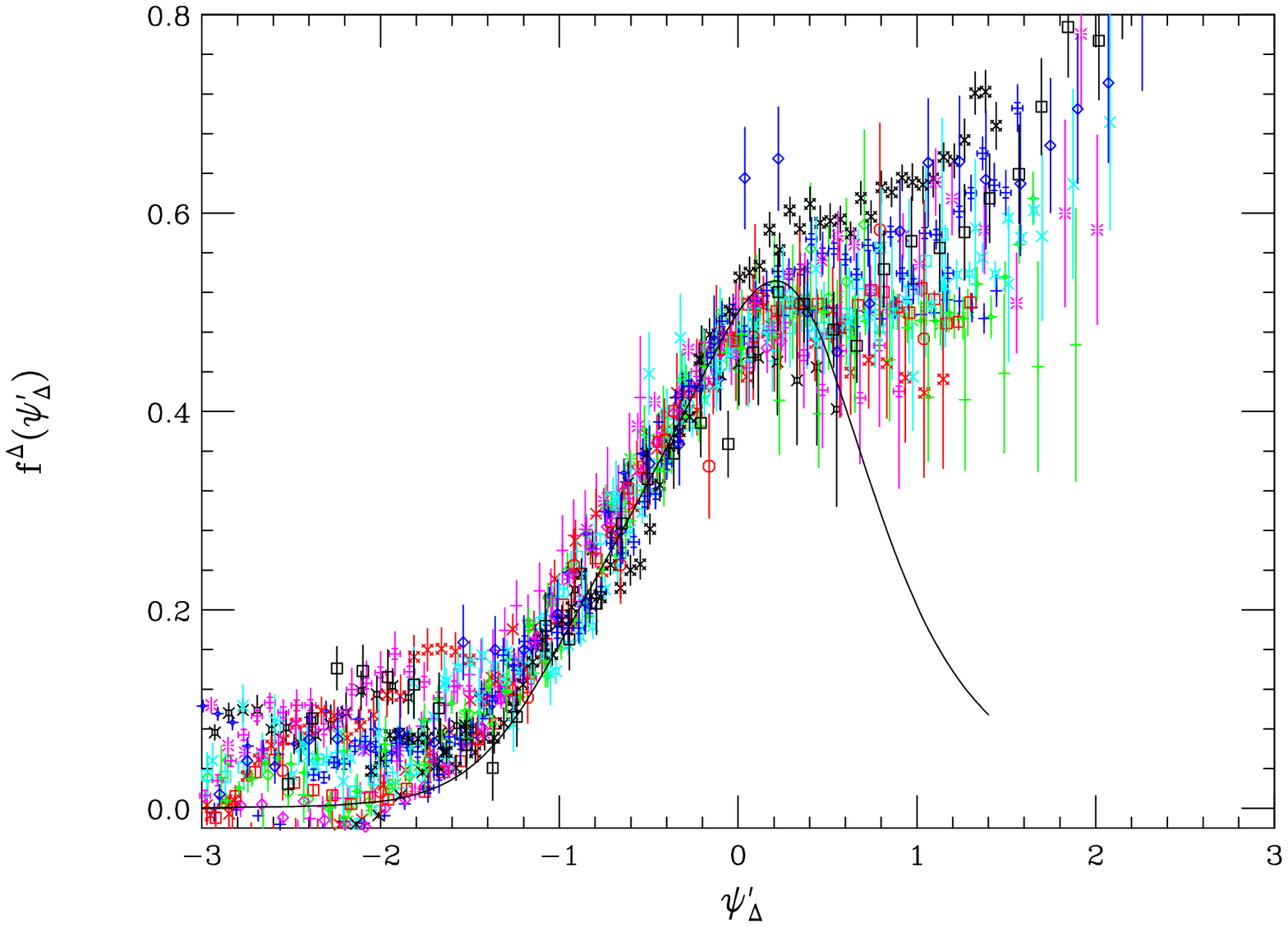}
%\parbox{13cm}
\vspace{-1cm}
{\caption{
Experimental QE and $\Delta$ 
superscaling functions versus the corresponding scaling
variables
% in the QE and $\Delta$ regions together with 
%phenomenological parameterizations of the data. 
}\label{fpsiplaf}} 
\end{center} 
\vspace{-1.cm}
\end{figure}
Let us now focus on the resonance region. 
In order to isolate the $\Delta$ contribution,
we subtract from the total experimental cross sections 
the quasi-elastic cross section recalculated 
using the universal 
$f^{QE}(\psi'_{QE})$ introduced above. 
That is, we remove the {\em impulsive} longitudinal and transverse 
contributions that arise from {\em elastic} $eN$ scattering, 
leaving (at least) MEC effects with their associated correlations and 
impulsive contributions arising from {\em inelastic} $eN$ scattering. 
The MEC effects can be ignored as they provide relatively 
small corrections~\cite{physrep}, and thus this yields a
response which is largely dominated by the $\Delta$, at least for energy losses
lower than the maximum of the $\Delta$ peak. 

Using the same procedure as in the QE peak, 
we can again reduce the left-over cross section by
dividing by the appropriate single-nucleon $N\to\Delta$ cross section
and display the result versus the scaling variable
$\psi'_\Delta$ defined in Eq.~(\ref{psi}).
In Fig.~1 (lower panel) 
we show the resulting $f^\Delta(\psi'_\Delta)$ extracted
from the high-quality world data for inclusive electron scattering from
$^{12}$C and $^{16}$O in the QE and $\Delta$ regions. 
These data span energies extending from 300 MeV to 4 GeV and scattering 
angles from 12 to 145 degrees, depending on the beam energy. 
As for $f^{QE}$, the experimental values of $f^\Delta$ have been parameterized
by a simple analytical function (solid line).
The data appear to scale reasonably well up to the peak of the $\Delta$
($\psi'_\Delta\cong 0$), although clearly for still 
higher excitation energies the scaling is broken by processes that are not 
well represented via $\Delta$-dominance 
(other resonances and, at the larger values of $q$, the tail of deep inelastic
scattering). 
There is also some excess at large negative $\psi'_\Delta$ which breaks the 
scaling to some degree, probably due to contributions from 
MEC and their associated correlations~\cite{physrep}.

The above analysis can now be checked by assembling all of
the pieces obtained via the scaling procedures to produce a total inclusive
cross section which can be compared with data. Several examples
of such a comparison are shown in Ref.~\cite{paper}. 
The answers are very
encouraging, and only for specific kinematics do we see deviations as large as
10--20\%, probably due to MEC and correlations. In particular the dip region,
usually hard to get from microscopic models, is nicely fitted in the present
context.
%
%\begin{figure}[hbt]
%\vspace{-0.6cm}
%\begin{center}
%  \includegraphics[scale=0.4,clip,angle=0]{fig2.ps}
  %\parbox{13cm}
%\vspace{-1.cm}
%{\caption{ Experimental $(e,e')$ cross section 
%for $^{12}$C,
%compared  with the calculated result obtained using superscaling (see text).
%$f^{QE}$ and
%      $f^\Delta$. The dashed curve is the QE contribution and the solid curve
%      is the total including the $\Delta$.
%  }\label{sigincl2}}
%\end{center} 
%\vspace{-1.cm}
%\end{figure}

We are then in a position to take a step in a new direction. Since the above
superscaling functions, upon being multiplied by the electromagnetic 
$N\to N$ and $N\to\Delta$ single-nucleon cross sections, reproduce quite well
the total nuclear electromagnetic cross section, 
we can just as well multiply by the
corresponding charge-changing weak interaction $N\to N$ and
$N\to\Delta$ single-nucleon cross sections to obtain predictions for neutrino
reactions in nuclei at similar kinematics. 
%Note that in the high-energy 
%domain considered here the dynamics in the axial sector, 
%not present in electron scattering, is not expected to differ significantly 
%from the one in the vector sector, since both V and A currents are 
%spin-dominated. This is not the case of course at low energies, where the
%scaling arguments cannot be applied.
The result for the reaction $^{12}$C$(\nu_\mu,\mu^-)$ is shown in Fig.~2 
versus the muon momentum $k'$ for typical kinematical conditions. 
Cross sections at different kinematics 
and for antineutrino scattering can be found in Ref.~\cite{paper}.
As discussed above, the predictions at momenta to the left of the $\Delta$ 
peak (excitations lying above the $\Delta$ region) are unreliable, since 
our scaling approach does not fully account for meson production, 
including resonances other than the $\Delta$, and DIS processes. 
\begin{figure}[hbt]
%\vspace{-0.8cm}
\begin{center}
\hspace*{1.5mm}
\includegraphics[scale=0.4,clip]{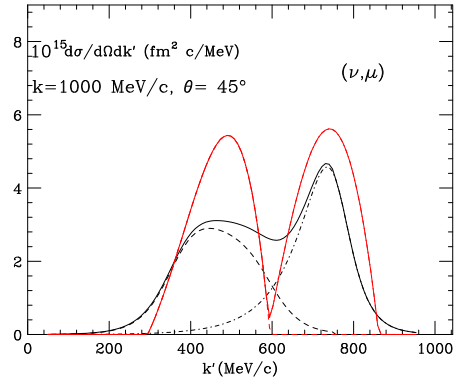}
%\parbox{13cm}
\vspace{-1.cm}
{\caption{%Charge-changing 
$\nu$-$^{12}$C cross section
%obtained with the superscaling method 
plotted versus the muon momentum $k'$. 
Dash-dotted: QE; dashed: $\Delta$; solid: total;  
heavier lines (red on-line): RFG.
}\label{neutoncarbon}} 
\end{center} 
\vspace{-1.cm}
\end{figure}
For comparison, in Fig.~2 we also show the cross section 
obtained using the RFG. 
Modeling based on mean-field theory where finals state interactions are
not included is very similar to the RFG 
result, since at the relatively high energies of interest the dynamical 
effects embodied in an effective mass are expected to be small. 
Likewise, relativized shell model predictions are close to the RFG 
predictions~\cite{Amaro:2005dn}
and, moreover, RPA correlations are expected to be relatively small for the 
high energies involved. Thus, the RFG predictions effectively represent a 
larger set of models. As can be seen in the figure, all therefore differ 
significantly from the scaling predictions. 

In conclusion, 
given the success of the scaling approach in studies of inclusive electron 
scattering for the kinematic region under study, we expect neutrino 
reaction cross sections also obtained using scaling ideas to be more robust 
than those based directly on existing models.

%\vspace{-0.3cm}

\end{document}